\documentstyle[psfig]{mn}
\newif\ifAMStwofonts
\newcommand{\lya}{Lyman-$\alpha$ }
\newcommand{\dla}{damped Lyman-$\alpha$ }

\ifoldfss

  \ifCUPmtlplainloaded \else
    \NewTextAlphabet{textbfit} {cmbxti10} {}
    \NewTextAlphabet{textbfss} {cmssbx10} {}
    \NewMathAlphabet{mathbfit} {cmbxti10} {} 
    \NewMathAlphabet{mathbfss} {cmssbx10} {} 
  \fi
  \ifAMStwofonts
    \ifCUPmtlplainloaded \else
      \NewSymbolFont{upmath} {eurm10}
      \NewSymbolFont{AMSa} {msam10}
      \NewMathSymbol{\upi}     {0}{upmath}{19}
      \NewMathSymbol{\umu}     {0}{upmath}{16}
      \NewMathSymbol{\upartial}{0}{upmath}{40}
      \NewMathSymbol{\leqslant}{3}{AMSa}{36}
      \NewMathSymbol{\geqslant}{3}{AMSa}{3E}

    \fi
  \fi
\fi 
\ifnfssone
  \newmathalphabet{\mathit}
  \addtoversion{normal}{\mathit}{cmr}{m}{it}
  \addtoversion{bold}{\mathit}{cmr}{bx}{it}

  \newmathalphabet{\mathbfit} 
  \addtoversion{normal}{\mathbfit}{cmr}{bx}{it}
  \addtoversion{bold}{\mathbfit}{cmr}{bx}{it}
  \newmathalphabet{\mathbfss} 
  \addtoversion{normal}{\mathbfss}{cmss}{bx}{n}
  \addtoversion{bold}{\mathbfss}{cmss}{bx}{n}
  \ifAMStwofonts
    \ifCUPmtlplainloaded \else
      \UseAMStwoboldmath
      \makeatletter
      \new@mathgroup\upmath@group
      \define@mathgroup\mv@normal\upmath@group{eur}{m}{n}
      \define@mathgroup\mv@bold\upmath@group{eur}{b}{n}
      \edef\UPM{\hexnumber\upmath@group}
      \new@mathgroup\amsa@group
      \define@mathgroup\mv@normal\amsa@group{msa}{m}{n}
      \define@mathgroup\mv@bold\amsa@group{msa}{m}{n}
      \edef\AMSa{\hexnumber\amsa@group}
      \makeatother
      \mathchardef\upi="0\UPM19
      \mathchardef\umu="0\UPM16
      \mathchardef\upartial="0\UPM40
      \mathchardef\leqslant="3\AMSa36
      \mathchardef\geqslant="3\AMSa3E

    \fi
  \fi
\fi 
\ifnfsstwo

  \DeclareMathAlphabet{\mathbfit}{OT1}{cmr}{bx}{it}
  \SetMathAlphabet\mathbfit{bold}{OT1}{cmr}{bx}{it}
  \DeclareMathAlphabet{\mathbfss}{OT1}{cmss}{bx}{n}
  \SetMathAlphabet\mathbfss{bold}{OT1}{cmss}{bx}{n}
  \ifAMStwofonts
    \ifCUPmtlplainloaded \else
      \DeclareSymbolFont{UPM}{U}{eur}{m}{n}
      \SetSymbolFont{UPM}{bold}{U}{eur}{b}{n}
      \DeclareSymbolFont{AMSa}{U}{msa}{m}{n}
      \DeclareMathSymbol{\upi}{0}{UPM}{"19}
      \DeclareMathSymbol{\umu}{0}{UPM}{"16}
      \DeclareMathSymbol{\upartial}{0}{UPM}{"40}
      \DeclareMathSymbol{\leqslant}{3}{AMSa}{"36}
      \DeclareMathSymbol{\geqslant}{3}{AMSa}{"3E}

    \fi
  \fi
 
\fi 

\ifCUPmtlplainloaded \else
  \ifAMStwofonts \else 
    \def\upi{\pi}
    \def\umu{\mu}
    \def\upartial{\partial}
  \fi
\fi
\title{Implications of 21cm observations for damped Ly-$\alpha$ systems}
\author[Jayaram N Chengalur $\&$ Nissim Kanekar  ]
       { Jayaram N Chengalur\thanks{chengalur@ncra.tifr.res.in} $\&$ 
 Nissim Kanekar\thanks{nissim@ncra.tifr.res.in}
 \\
        National Centre for Radio Astrophysics, Tata Institute of Fundamental
        Research, Pune - 411007, India}
\pagerange{\pageref{firstpage}--\pageref{lastpage}}
\begin{document}
\maketitle

\begin{abstract}

	We present Giant Metrewave Radio Telescope HI 21cm absorption 
observations, of candidate and confirmed \dla systems. The derived 
spin temperatures T$_{\rm spin}$ are high, in all cases, $\sim 1000$~K 
or higher. We have also collated from the published literature a list 
of damped absorbers for which 21cm observations exist, and discuss 
the implications of the observations for the nature of these systems.

	A cross-comparison of the HI 21cm profiles (which trace the
{\it cold} gas) with the low ionization metal profiles (which trace
{\it all} the neutral HI, both cold and warm) shows that in all cases for
which both spectra are available, the 21cm absorption coincides in velocity
with the deepest metal line feature. This is consistent with models 
in which the deep metal line features arise from discrete clouds but 
not with models where the deepest features are the result of velocity
crowding.

	We also find that the typical derived spin temperatures of 
\dla systems are considerably higher than those in the Milky Way 
or nearby spiral galaxies. The only exceptions are systems which
are known to be associated with the disks of spirals; these
{\it do}, in fact, show low spin temperature. In a multi-phase medium
the derived spin temperature is a weighted average of the temperatures
of the individual phases. High apparent T$_{\rm spin}$ values are 
hence to be expected from small, low metallicity objects since 
these objects should (as per existing theoretical models of the 
formation of a multi-phase ISM in the Milky Way and  high redshift
proto-galaxies) have a lower fraction of the cold phase in their 
ISM as compared to large galaxies. The high  T$_{\rm spin}$ 
is hence consistent with the observed low metallicities of \dla 
absorbers as well as with recent findings that damped absorption 
is associated with a variety of galaxy types (as opposed to being 
confined to the disks of large spirals).

	Finally, although the number of systems for which observations
are available is small, we suggest that the following trend may be 
identified : at low redshift, \dla absorption arises from a range of 
systems, including spiral galaxy disks, while, at high redshift, 
absorption occurs predominantly in smaller systems.
\end{abstract}


\label{firstpage}
\section{Introduction.}
\label{sec:introduction}
	A gas cloud with a column density $N_{\rm HI} > 10^{20}$ atoms/cm$^{-2}$
produces a broad Lyman$-\alpha$ absorption profile with characteristic 
Lorentzian damping wings. Such systems can hence be easily identified 
from even modest resolution spectra of distant quasars. Although these high
column density absorbers are extremely rare compared to the ubiquitous
\lya forest lines, they nonetheless make the dominant contribution to 
the observed {\it neutral } hydrogen at high redshifts.

	At $z=0$, the cross section for \dla absorption is dominated
by the disks of spiral galaxies (Rao \& Briggs 1993). It is thus plausible
that the damped systems seen at high redshift are either spiral galaxies
or the precursors of spiral galaxies. In support of this hypothesis, large 
surveys of these objects have established that the mass density of 
neutral gas in \dla systems at $z\sim 3$ is comparable to the mass 
density of stars in luminous galaxies at $z\sim 0$ (eg. Wolfe 1988, 
Lanzetta et al. 1991, Storrie-Lombardi et al. 1996). Further, the mass 
density in neutral gas today is far smaller than that at high redshift. 
The evolution of $\Omega_{\rm HI}$ with redshift thus matches the pattern 
expected from gas depletion due to star formation. 

	While the motivation for the original \dla surveys was to 
find disk galaxies at high redshift (Wolfe et al. 1986), the nature of 
the systems in which this absorption arises remains as yet 
controversial.  Absorption line surveys principally establish the 
probability of finding an absorber along a random line of sight. This 
probability is proportional to the product of the volume number density 
of the absorbers and their typical cross-section; neither of the two
is separately constrained. The observed frequency of occurrence of 
\dla lines in QSO spectra at high $z$ is about 5 times larger than that
expected from a population in which both the HI disk cross-section and 
the galaxy number density remain constant (Wolfe et al. 1986, Smith,
Cohen \& Bradley 1986, Lanzetta et al. 1991). The nominal conclusion is 
that, if damped absorption traces the precursors of spiral galaxies, then 
either the typical galaxy disk was somewhat larger in the past than it 
is now, or there were considerably more galaxies in the past than 
there are today. 

	Prochaska \& Wolfe (1997,1998) attempted to settle the issue of 
the nature of damped absorbers in a purely observational manner, using 
high resolution spectra of low-ion metal transitions, to investigate the 
velocity structure of 31 systems. These low-ionization metals co-exist 
with neutral HI; their absorption profiles, however, are unsaturated and
hence trace the kinematics of the neutral gas along the narrow line
of sight to the background QSO. The authors found that the
metal line profiles are systematically asymmetric (in that the
deepest absorption preferentially occurs at the profile edge).
Such asymmetries arise naturally in absorption from a spinning 
disk due to two effects (i) a ``radial effect'', where the strongest 
absorption occurs at the tangent point, i.e. due to velocity crowding, 
and (ii) a ``perpendicular effect'' in which 
the strongest absorption takes place in the midplane, if the line-of-sight 
intersects the midplane more than a few scale heights away from the 
major axis. The metal line profiles thus appear to support the hypothesis
that \dla absorption arises primarily in spiral galaxy disks. This is 
controversial, however, since the observed asymmetries can also be 
explained by merging proto-galactic clumps in hierarchical models 
(Haehnelt et al. 1998) and even by randomly moving clouds in a 
spherical halo (McDonald \& Miralda-Escude 1999). 

	At low and intermediate redshifts, there have been a number of 
ground-based, as well as HST, observations to try and directly image 
the systems responsible for the absorption (see, for example, le Brun 
et al. 1997). From these observations, it appears that, contrary to 
predictions based on the HI content of $z=0$ galaxies, \dla absorption 
is {\it not} associated predominantly with bright ($L > L_{\star}$) 
spiral galaxies, but appears to arise in galaxies spanning a wide 
range of morphological types. Interestingly, even in quasar-galaxy 
pairs at very low redshift, systems which show HI absorption appear 
to be predominantly disturbed or interacting galaxies (Carilli \& van 
Gorkom 1992).

	If the quasar against which the \lya absorption is seen is
radio loud, one can also attempt to detect redshifted 21cm
absorption from the \dla system (eg. Wolfe \& Davis 1979,
Wolfe, Briggs \& Jauncey 1981, Wolfe et al. 1985,
de Bruyn et al. 1996, Carilli et al. 1996, Lane et al. 1998,
Chengalur \& Kanekar 1999). Observing the  21cm line yields a
number of interesting physical parameters. Firstly, since the 21cm
line is not saturated, the absorption profile reflects the kinematics
of the cold gas in the system. Further, (under suitable assumptions)
the 21cm optical depth can be used along with the total HI column density 
(obtained from the \dla profile) to yield the average spin temperature of the 
gas. Under a variety of astrophysical conditions (Field 1958), 
the spin temperature of the 21cm transition is identical to the
kinetic temperature of the gas; it is hence a useful diagnostic.
Finally, if the background source is extended, one can attempt 
to determine the transverse extent of the absorbing gas. This 
cannot be done using UV absorption lines since the background 
quasar is unresolved at these wavelengths.

	We present, in this paper, Giant Metrewave Radio Telescope (GMRT) 
HI 21cm observations of candidate and confirmed low redshift \dla systems. 
We have also collated from the published literature earlier 21cm studies of 
a set of such objects, and discuss the implications of these observations 
for the nature of the absorbers.

\section{Observations \& Results}
\label{sec:Obs+Res}

	The observations were carried out in April and October 1999 as 
part of the commissioning of the first sideband of the 30 station 
correlator at the GMRT. The number of antennas available was typically 
$\sim 10$ because of various on-going debugging and maintenance activities. 
Both linear polarizations were observed. The GMRT correlator is an 
FX correlator which provides a fixed number (128) of spectral channels 
(per sideband) over a bandwidth that can be varied from 16~MHz to 64~kHz. 
The current observations used bandwidths of 1 and 2~MHz and no online 
spectral windowing. 

\noindent {\bf 3C446 \& PKS 1451-375} : Both PKS~1451-375 and 
3C~446 are listed as candidate damped Lyman-$\alpha$ absorbers on the
basis  of their IUE spectra (Lanzetta et al. 1995). The present  
observations used a total bandwidth of 2 MHz, centred at the frequency
of the optically determined redshift (See Table 1 for details).
No statistically significant absorption was detected in either case. The 
RMS noise levels in the spectra after a single Hanning smoothing (i.e. an 
effective velocity resolution of $\sim 9$ km s$^{-1}$) were 7 mJy 
for 3C446 and 2 mJy for 1451-375, with on-source times of $\sim$ 1 hour and 
$\sim$ 100 minutes on 3C446 and 1451-375 respectively.

\begin{table*}
\label{tab:OBS}
\caption{Observing details}
\begin{tabular}{|r|r|r|r|r|r|}
Source& $z_{abs}$ & N$_{\rm HI}$ & Resolution& $\tau$ & T$_s$  \\
& &cm$^{-2}$ &km s$^{-1}$ & & K\\
&&&&& \\
PKS 1451-375    &0.2761       &1.3e20$^a$    & 4.2    & $<$ 0.006  & $>$ 1400     \\
PKS 1127-145    &0.3127       &5.1e21$^b$    & 2.17   & 0.092      & 910          \\
3C446           &0.4842       &6.3e20$^a$    & 4.8    & $<$ 0.02   & $>$ 1600     \\
&&&&& \\
\end{tabular}

${}$~Limits are 3$\sigma$.\\
${}^a$~Lanzetta et al. (1995),${}^b$~Rao \& Turnshek (1999) 
\end{table*}

	3C~446 has complex structure at milli-arcsecond scales (Simon 
et al. 1985) with a total flux of $\sim$ 1.1 Jy within the central 
10 milli-arcseconds. If this central compact region is completely 
covered by the absorbing gas (as seems likely), the 3$\sigma$ lower 
limit to the spin temperature is $\sim$ 1600 K (see section~\ref{sec:dis} 
for a discussion on the derivation of the spin temperature). In the case of 
1451-375, the milli-arcsecond structure is not known. It does, however, have an 
unresolved sub-arcsecond sized core, with an inverted spectrum and a 
total flux of 1.1 Jy at 1.4~GHz, If one assumes that the absorber 
completely covers this central core, the 3$\sigma$ lower limit to the 
spin temperature is $\sim$ 1400 K. 

\begin{figure}
\centering
\psfig{file=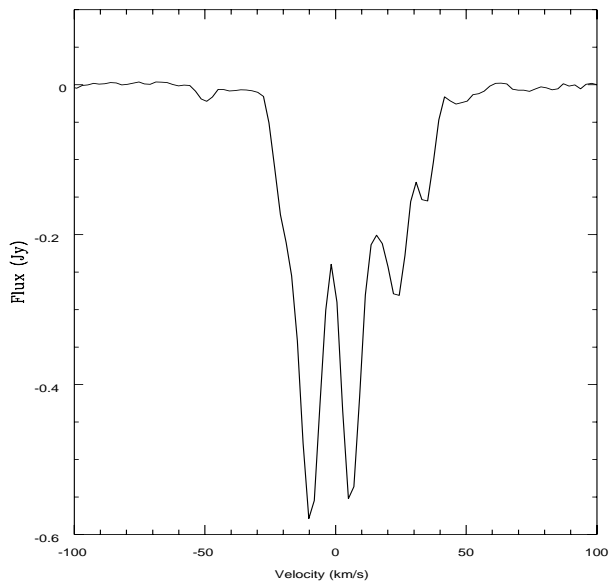,width=3.25truein,height=3.25truein}
\caption{ GMRT HI spectrum towards PKS~1127-145. The velocity axis is 
centred at $z = 0.3127$.}
\label{fig:b1127}
\end{figure}

\noindent {\bf PKS 1127-145} : The 21cm absorption line system towards 
PKS~1127-145  was discovered by Lane et al. (1998), using the WSRT. The
observations, however, did not resolve out the line and were also 
corrupted by low-level interference close to the line location. 
Figure~\ref{fig:b1127} shows a higher resolution and higher sensitivity 
GMRT spectrum. The channel spacing is $\sim$ 2~km/s and the RMS noise 
after a single Hanning smoothing is $\sim$ 3.5 mJy. The broad feature 
detected  by Lane et al. (1998) breaks up into a number of narrow components.
There is also an entirely new component, $\sim$ -50 km s$^{-1}$ 
from $z = 0.3127$, not detected in the WSRT data. This feature is present 
in more than one observing run, with the correct Doppler shift. The
total velocity range over which the absorption is seen is $\sim 120$~km/s. 

	VLBI observations at 1667 MHz of PKS~1127-145 (Bondi et 
al. 1996)  have shown that it has a core-jet structure, with a total size 
less than $\sim$ 20 milli-arcseconds. This small size implies that the 
covering factor of  the absorbing system is likely to be close to unity.
Assuming a covering factor of unity results in a spin temperature 
$T_{\rm s}$ of $ \sim 910 \pm 160$~K, consistent within the error bars 
with the value of $1000 \pm 200$~K obtained by Lane et al. (1998).

\section{Discussion}
\label{sec:dis}

	We were able to collate from the published literature eighteen
21cm observations of \dla absorbers, including cases where only upper limits 
have been obtained on the optical depth. This is probably {\it not} an 
unbiased  sample since it is unlikely that all non-detections have been 
published. The relevant parameters of the sample are summarised in 
Table 2. Note that although 3C~446 and PKS~1451-375 are included
in Table 2 and Figure 2, their status as bonafide \dla systems remains 
to be confirmed.
	
	The optical depth in the 21cm transition is a function of both
the column density, $N_{\rm HI}$, of the gas as well as its spin 
temperature, $T_s$. For an optically thin, homogenous cloud the exact
relation is :
\begin{equation}
\label{eqn:Tspin}
	N_{\rm HI} = { 1.823\times10^{18} T_s \over f} \int \tau_{21} dV
\end{equation}
where $\tau_{21}$ is the 21cm optical depth, $N_{\rm HI}$ is the neutral 
hydrogen column density and $f$ is the covering factor; $T_s$ is 
in K and $dV$ in km s$^{-1}$. For an inhomogeneous absorber, the 
temperature $T_s$ in equation (\ref{eqn:Tspin}) should be 
replaced by the column density weighted harmonic mean of the 
temperatures of the different components (assuming all components 
are optically thin). Thus, for example, if one has an absorber in
which 50\% of the gas is at 80~K and 50\% at 8000~K, the measured
spin temperature would be $\sim 160$~K, while if 10\% is at 80~K
and 90\% at 8000~K, the measured spin temperature would be 
$\sim 730$~K.

	In the case of damped systems, the column density 
$N_{\rm HI}$ is known from the \dla profile; this can be used in 
conjunction with the measured 21cm optical depth to obtain the spin 
temperature of the absorbing gas, if the covering factor $f$ is known. 
Many of the background sources listed in Table 2 are 
extremely compact making it likely that the covering factor is close 
to unity. Further, as discussed in more detail in Carilli et al. 
(1996), although the covering factor $f$ is sometimes poorly constrained,
it is unlikely to lead to a systematic bias for the systems discussed
here. We also emphasise that the sample of Table 2 does 
not contain unpublished non-detections of absorption; since $T_s$ 
and $\tau_{21}$ are inversely related, the bias in the sample (if it 
exists) is thus towards {\it lower} $T_s$ values. As we will see below,
\dla systems tend to have $T_s$ values higher than those observed in local
spiral galaxies; the results from our sample are hence conservative since 
the bias tends to reduce this discrepancy.

\subsection{Kinematics of damped systems}
\label{ssec:Kinematics}

	A more fundamental consequence of the inverse relationship
between $\tau_{21}$ and $T_s$ is that 21cm absorption typically traces the 
cold gas ($T_s \sim 100$ K) in the absorber. For example, in our galaxy,
the warm neutral medium (WNM) has been detected in absorption only towards
the very bright radio source Cygnus~A (Carilli et al. 1998), while, in 
extra-galactic systems, there has been only one claim of 21cm absorption 
that could arise from warm ($T_s \sim 5000$~K) gas (Lane et al. 1999). 
Low ionization metal lines, on the other hand, trace both the warm and 
the cold neutral gas. A comparison between 21cm profiles and the profiles 
of metal lines can thus be used to glean further information on the 
phase structure and kinematics of the absorbing  system. Unfortunately, 
there are only four known 21cm absorbers for which high resolution 
($\sim$ 8 km s$^{-1}$) optical spectra are available. These are 
discussed individually below. \\

\begin{table*}
\label{basicdata}
\caption{ Damped Ly-$\alpha$ Systems with 21cm Observations}
  \begin{tabular}{@{}lllllrll}
Name & $z_{abs}$ & N$_{\rm HI}$ & No. of & ${\Delta V_{21}}^{  a}$ & T$_{\rm s}^{  b}$ & Absorber ID& Ref\\
&&cm$^{-2}$& Components& km s$^{-1}$&K&& \\    
&&&&&&& \\    
0738+313 (A)    &0.09123      &1.5e21    &3   &25          &825           & Dwarf ?         & 1,2 \\
0738+313 (B)    &0.2212       &7.9e20    &1   &16          &1120          & Dwarf ?         & 1-3 \\
PKS~1413+135    &0.24671      &2.0e22    &1   &39          &150           & Spiral         & 4  \\
PKS~1451-375$^c$&0.2761       &1.3e20    &-   &-           &$>$ 1400      & -              & 5,6 \\
PKS~1127-145    &0.3127       &5.1e21    &7   &120         &910           & Group          & 3,6 \\
PKS~1229-021    &0.39498      &5.6e20    &2   &180         &170           & Spiral ?        & 7-10 \\
3C196           &0.4366       &6.3e20    &3   &250         &-             & Barred Spiral  & 11-13 \\
3C446$^c$       &0.4842       &6.3e20    &-   &-           &$>$ 1600      & -              & 5,6 \\
A0~0235+164     &0.524        &5.0e21    &5   &125         &170           & Spiral         & 14-17 \\
3C286           &0.69215      &2.0e21    &1   &18          &1200          & LSB            & 9,18-20\\
PKS~0454+039    &0.8596       &5.0e20    &-   &-           &$>$870        & Galaxy         & 9,21,22 \\
MC3~1331+170    &1.77636      &7.6e21    &1   &25          &770           & -              & 23-25 \\
PKS~1157+014    &1.944        &6.3e21    &1   &60          &470           & Group          & 26-28 \\
PKS~0458-02     &2.03945      &8.0e21    &2   &30          &594           & -              & 29,30 \\
PKS~0528-2505   &2.8110       &2.2e21    &-   &-           &$>$ 730       & Group          & 31-33 \\
2342+342        &2.9084       &2.0e21    &-   &-           &$>$ 1800      & -              & 31,34 \\
PKS~0336-017    &3.0619       &1.5e21    &-   &-           &$>$ 2500      & -              & 31,35\\
PKS~0201+113        &3.3875       &2.5e21    &-   &-           &$>$ 5600      & -              & 34,36-38 \\
&&&&&&& \\
\end{tabular}

${}^a$~Since many spectra have multiple-components and are not well 
fit by a single Gaussian, the width quoted here is the entire velocity 
range over which absorption is seen. 
${}^b$~The spin temperature (or its limits) quoted here has been 
consistently (re)computed from the published data using equation
(\ref{eqn:Tspin}), and the latest available values for all parameters in the 
equation. In certain instances, this value may differ slightly from 
that quoted by the authors in the original reference. The only exception
is PKS~0201+113, for which multiple and quite different values are quoted in
the literature. We use the value from Kanekar \& Chengalur (1997).
${}^c$~Candidate \dla systems, based on IUE spectra.

${}^{1}$~Rao \& Turnshek (1998),
${}^{2}$~Chengalur \& Kanekar (1999), 
${}^{3}$~Lane et al. (1998), 
${}^{4}$~Carilli et al. (1992),
${}^{5}$~Lanzetta et al. (1995),
${}^{6}$~This paper,
${}^{7}$~Brown \& Spencer (1979),
${}^{8}$~Briggs (1998),
${}^{9}$~Boiss\'{e} et al. (1998),
${}^{10}$~le Brun et al. (1997),
${}^{11}$~Brown \& Mitchell (1983),
${}^{12}$~de Bruyn et al. (2000),
${}^{13}$~Cohen et al. (1996),
${}^{14}$~Roberts et al. (1976),
${}^{15}$~Wolfe et al. 1982,
${}^{16}$~Burbidge et al. (1996),
${}^{17}$~Cohen et al. (1999),
${}^{18}$~Brown \& Roberts (1973),
${}^{19}$~Davis \& May (1978), 
${}^{20}$~Steidel et al. (1994),
${}^{21}$~Steidel et al. (1995),
${}^{22}$~Briggs \& Wolfe (1983),
${}^{23}$~Wolfe \& Davis (1979), 
${}^{24}$~Wolfe et al. (1976).
${}^{25}$~Carswell et al. 1975
${}^{26}$~Wolfe et al. (1981),
${}^{27}$~Briggs et al. (1984),
${}^{28}$~Fynbo et al. (1999),
${}^{29}$~Wolfe et al. (1985),
${}^{30}$~Briggs et al. (1989),
${}^{31}$~Carilli et al. (1996),
${}^{32}$~M{\o}ller \& Warren (1993),
${}^{33}$~M{\o}ller \& Warren (1998),
${}^{34}$~White et al. (1993),
${}^{35}$~Lu et al. (1993),
${}^{36}$~de Bruyn et al. (1996),
${}^{37}$~Briggs et al. (1997),
${}^{38}$~Kanekar \& Chengalur (1997).
\end{table*}

\noindent 1. 1229-021 ($z_{abs} = 0.3950$) The metal lines of this 
system were studied by  Lanzetta \& Bowen (1992), using the AAT with a
velocity resolution of $7$~km s$^{-1}$. The 21cm spectrum (Brown \& Spencer 1979;
Briggs 1998) shows a narrow component centered at the redshift of 
the deepest part of the metal line profile. The metal lines show two more
components at $+100$~km s$^{-1}$ and $+200$~km s$^{-1}$, relative to 
$z = 0.395$. While the 21cm profile has a broad shallow component with 
a total width between nulls of $180$~km s$^{-1}$ there is no absorption 
associated with the $+200$~km s$^{-1}$ metal line component nor a 
separate discrete component at $+100$~km s$^{-1}$.

\noindent 2. 3C286 ($z_{abs}$ = 0.69215) The velocity structure of the 
damped system at $z \sim 0.692$ towards 3C286 is exceedingly simple in 
both the 21cm and low-ion (FeII, CaI) transitions. Each consist of a 
single narrow component with b values of $5.2$~km/s for the 21cm 
line (Davis \& May 1978) and $6.5$~km/s for the FeII line (Meyer \& York 
1992). The metal lines occur at $z = 0.69218 \pm 0.00007$ while the 21cm 
line is at $z = 0.692154 \pm 0.000002$; thus, the locations are coincident 
in velocity, within the error bars.

\noindent 3. 1331+170 ($z_{abs} = 1.7764$) The SiII 1808 line detected 
from this system (resolution $\sim 6.6$~km/s) is clearly asymmetric 
(Prochaska \& Wolfe 1997), with its deepest feature at 
$z = 1.77636 \pm 0.00006$ . The 21cm profile (Wolfe \& Davis, 1979) has
a single component at $z = 1.77642 \pm 0.0001$, the deepest part of the
SiII line coincides with the 21cm line, again within the error bars. 
However, once again there are discrete components in the SiII spectrum 
which have no counterparts in 21cm absorption.

\noindent 4. 0458-020 ($z_{abs} = 2.039$) The 21cm line (Wolfe et al. 1985) 
has two clear components, at $z = 2.03937$ and $z = 2.03955$. The two deepest
features of the CrII~2056 profile (Prochaska \& Wolfe 1997) coincide with 
the 21cm features (within the error bars); however, the deeper CrII 
feature corresponds to the weaker 21cm feature (see Figure 2, in 
Prochaska (1999), for an overlay of the two profiles). There are also 
several weaker CrII features which have no corresponding features in 
the 21cm spectrum. 

	Four systems are clearly too small a sample to draw any far-reaching 
conclusions regarding the general kinematics of damped absorbers. It is, 
however, revealing that the strongest metal line absorption coincides in 
velocity with a 21cm absorbing component, in every case for which 
high-resolution optical spectra are available. This points to a physical 
coincidence between the cold gas responsible for the 21cm absorption and 
the (not necessarily cold) gas giving rise to the deepest features in 
the low-ion profiles. This is consistent with any model for which the 
absorption arises in discrete clouds, but not with models for which the 
deep features in the low-ion profiles are the result of velocity crowding.

	The other noteworthy feature is that the low ionization metal lines
are in general much broader than the 21cm profiles, and often contain
discrete velocity components which have no analogues in the 21cm spectrum.
For example, if one assumes that the neutral hydrogen column density scales 
with the depth of the low ionization lines (which is {\it not} necessarily 
a good assumption), the noise levels on the 21cm spectrum towards 
0458-020 (Wolfe et al. 1985) are sufficiently low to detect 21cm absorption 
corresponding to the feature at $v \sim 50$ km s$^{-1}$ in the CrII 
spectrum of Prochaska \& Wolfe (1997) (see Figure 2 of Prochaska (1999)). 
Much as in our own ISM, the components with no corresponding 21cm absorption 
probably arise in warm gas which contributes to the total 
$N_{\rm HI}$ measured in the Ly$-\alpha$ line, but whose spin temperature
is too high for it to be seen in 21cm absorption. It is possible, of course,
that the above UV absorption arises from a number of small cold clouds, 
with high inter-cloud velocity dispersion, with the column density of each 
individual cloud being too low to be detected in the 21cm line. This is, 
however, an unlikely scenario since the clouds would require a column 
density of $\sim 10^{19}$ cm$^{-2}$ for self-shielding and column densities
as low as $\sim 10^{18}$ of {\it cold} HI can typically be detected in
21cm absorption. 

\begin{figure}
\centering
\psfig{file=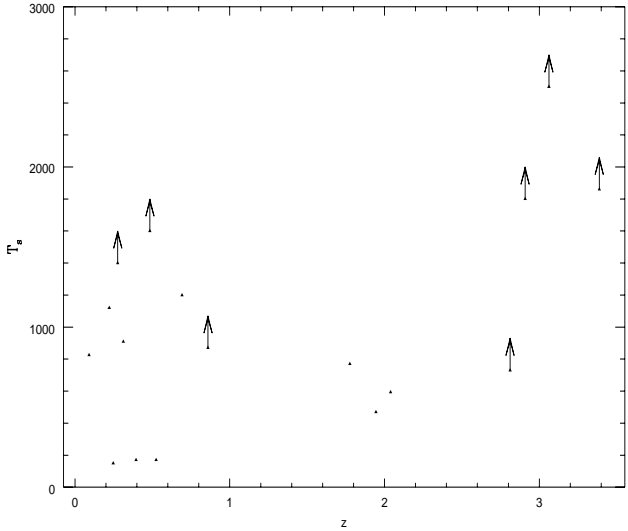,width=3.5truein,height=3.0truein,angle=-90}
\caption{ Plot of $T_s$ for all damped systems for which published 21cm 
absorption measurements exist. Upper limits are shown by upward pointing
arrows. Absorbers identified as spiral galaxies are shown as hollow
squares.}
\label{fig:Tspin}
\end{figure}

\subsection{The spin temperature of damped absorbers}
\label{subsec:Tspin}

	Figure~\ref{fig:Tspin} shows a plot of spin temperature v/s 
redshift for 17 systems from Table 2 (excluding 3C196, for which the 
background source is extended, making the column density uncertain). 
It can be clearly seen that the majority of systems tend 
to have $T_s$ values much larger than 500~K. For comparison, spin 
temperatures in the disc of the Galaxy or Andromeda are typically 
between 100 and 200 K (Braun \& Walterbos 1992; Braun 1997).
Similarly, Carilli \& van Gorkom (1992) find that when the 
line-of-sight passes through the disc of a galaxy, the spin temperatures
derived from 21cm emission-absorption arising in nearby quasar-galaxy 
pairs are also low, i.e. within 1 $\sigma$ of 300 K, (although the error 
bars are admittedly large).  Further, as discussed by 
Carilli et al.~(1996), the difference in $T_s$ values between damped 
systems and local discs is probably not due to the different methods 
used for the measurements in the two cases.

	Interestingly enough, however, {\it all} the absorbers in 
Table~2  which have been identified as spiral galaxies either have 
low $(<$ 200 K) spin temperatures or, where the column density is not 
independently known (in the case of the extended source, 3C196), have 
absorption profiles consistent with low $T_s$ values for moderate values 
of column density (de Bruyn, Briggs \& Vermeulen 2000). It thus appears 
that low spin temperatures {\it do} occur in damped Lyman-$\alpha$ systems,
when the line-of-sight passes through the disk of a spiral galaxy. This 
indicates that the discrepancy in spin temperature between the majority
of damped absorbers and local spirals is a real one, which may well arise
from a true difference in the physical characteristics of the two classes
of systems. The likely explanation for the high spin temperatures in 
damped systems is (as suggested earlier, eg. Carilli et al. 1996) that 
their fractional content of warm gas is larger than that of local spirals.

	 In the Milky Way, neutral gas can exist in pressure equilibrium 
in two stable phases, the warm neutral medium (WNM) at T~$\sim 10^4$~K and the
cold neutral medium (CNM) at T $\sim 60$~K. At high pressures (such as those
obtained in the midplane  of the Galaxy), only the cold phase exists, while,
at low pressures (as obtained at  large scale heights, far away from the 
midplane), only the warm phase is found.  At intermediate pressures, the two
phases co-exist in pressure equilibrium. Detailed models of heating and 
cooling processes in a two-phase neutral medium (Wolfire et al. 1995) have 
shown that the pressure required to produce the  cold phase increases with
a decrease in metallicity. The cold phase should hence  be rare in smaller 
galaxies which have both lower midplane pressures and lower metallicities.
Since the fractional contribution of the CNM to the total column density 
would be smaller in such systems, they would (as detailed earlier) have 
higher average $T_s$ values than spiral galaxies.

	The metallicity evolution of \dla systems is currently controversial.
Damped absorbers typically have low metallicity ($\sim 0.1$ solar) and 
there is little direct evidence for evolution of metallicity with redshift
(Pettini et al. 1999). However Savaglio \& Panagia (1999) find that if one 
assumes, and fits for, depletion using depletion patterns observed in 
the Milky Way, the metallicity does appear to increase with decreasing 
redshift. If the absorbers were indeed spirals, one would have expected 
to see metallicity evolution. If, on the other hand, damped systems 
typically arise in smaller galaxies, a consistent picture can be seen 
to emerge. Smaller galaxies have lower levels of star formation and are 
hence generally metal poor. Further, their midplane pressures are lower 
than those of large galaxies, making it doubly difficult to create the 
conditions for the formation of the CNM. Thus, both the low metallicity 
and the high spin temperature of damped systems can be consistently 
explained if damped Lyman-$\alpha$ absorption occurs predominantly 
in small rather than large galaxies.  Finally, a detailed analysis
of the formation of a multi-phase medium in proto-galaxies (Spaans \& Norman 
1997, Spaans \& Carollo 1997)  indicates that the epoch of formation of 
a multi-phase ISM is a strong function of the galactic mass; while galaxies 
with baryonic mass $\sim 5 \times 10^{11}$~M$_\odot$ form the cold phase
of the ISM by redshifts of $\sim 3$, dwarf galaxies with mass 
$\sim 3\times 10^9$~M$_\odot$ typically form the cold phase only
by redshfits $\sim 1$.

	It can be seen from Figure~\ref{fig:Tspin} that low spin 
temperatures are not obtained at high redshift ($z \ga
2$), while both low and high $T_s$ values are found at low 
redshift. Further, Table~\ref{basicdata} indicates that large velocity
widths ($\Delta V_{21} > 100$ km s$^{-1}$) are not found for $z 
\ga 2$, while both large and small widths are seen at 
low redshift. This is qualitatively consistent with what is expected
in hierarchical merging models, (eg. Kauffmann 1996). Interestingly,
$1127-145$ is the only system which has both a large 21cm velocity 
spread and a high temperature; all other systems with  large velocity 
widths are identified with low $T_s$ spirals. There are, 
however, at least three galaxies at the redshift of the \dla absorber 
(Lane et al. 1998). It is possible that they are interacting and that some of 
the HI in this system is associated with tidal features and has a high 
spin temperature. 

	Hence, although the numbers are small, the general trends in 
$T_s$ and $\Delta V_{21}$ are consistent with scenarious in which 
\dla absorbers are typically small systems at high redshifts, while at 
low redshifts, they are a composite population including both 
spiral galaxies as well as smaller systems. This would be consistent both 
with hierarchical models of galaxy formation and narrow band imaging 
observations (Fynbo et al. 1999), which show that high redshift \dla 
systems often have several Ly-$\alpha$ emitting objects at the 
redshift of the absorber.

\noindent {\bf Acknowledgments} The observations presented in this paper would 
not have been possible without the many years of dedicated effort put in 
by the GMRT staff to build the telescope. We are also grateful to the
referee for useful comments.



\end{document}